\documentclass[aps,pre,twocolumn,nolongbibliography,amsmath,amssymb,amsfonts,superscriptaddress,nofootinbib]{revtex4-2}

\usepackage{amsmath}
\usepackage{amsfonts}
\usepackage{comment}
\usepackage{graphicx} 
\usepackage{xcolor}
\usepackage{multirow}
\usepackage{hyperref}

\usepackage{times}

\newcommand{\rr}{\boldsymbol{r}}

\date{\today}

\begin{document}

\title{Exponents and front fluctuations in the quenched Kardar-Parisi-Zhang universality class of one- and two- dimensional interfaces}
\author{A. Tajuelo-Valbuena}
\affiliation{Departamento de F\'{\i}sica, Universidad de Extremadura, 06006 Badajoz, Spain}
\author{J. Trujillo-Mulero}
\altaffiliation{Current address: Departamento de Física Teórica de la Materia Condensada, Universidad Autónoma de Madrid, 28049 Madrid, Spain}
\affiliation{Departamento de F\'{\i}sica, Universidad de Extremadura, 06006 Badajoz, Spain}
\author{J. J. Meléndez}
\affiliation{Departamento de F\'{\i}sica, Universidad de Extremadura, 06006 Badajoz, Spain}
\affiliation{Instituto de Computaci\'on Cient\'{\i}fica Avanzada de Extremadura (ICCAEx), Universidad de Extremadura, 06006 Badajoz, Spain}
\author{R. Cuerno}
\affiliation{Universidad Carlos III de Madrid, Departamento de Matem\'aticas and Grupo Interdisciplinar de Sistemas Complejos (GISC), Avenida de la Universidad 30, 28911 Legan\'es, Spain}
\author{J. J. Ruiz-Lorenzo}
\affiliation{Departamento de F\'{\i}sica, Universidad de Extremadura, 06006 Badajoz, Spain}
\affiliation{Instituto de Computaci\'on Cient\'{\i}fica Avanzada de Extremadura (ICCAEx), Universidad de Extremadura, 06006 Badajoz, Spain}

\begin{abstract}
We have simulated an automaton version of the quenched Kardar-Parisi-Zhang (qKPZ) equation in one and two dimensions in order to study the scaling properties of the interface at the depinning transition. Specifically, the $\alpha$, $\beta$, $\theta$, and $\delta$ critical exponents characterizing the surface kinetic roughening and depinning behaviors have been directly computed from the simulations. In addition, by studying the height-difference correlation function in real space, we have also been able to directly compute the dynamic correlation length and its associated dynamic critical exponent $z$. The full sets of scaling exponents are largely compatible with those of the Directed Percolation Depinning universality class for one and two dimensional interfaces. Furthermore, we have computed numerically the probability density function (PDF) of the front fluctuations in the growth regime, finding its asymptotic form in one and two dimensions. While the PDF features strongly non-Gaussian skewness and kurtosis values, it also differs from the PDF of the KPZ equation with time-dependent noise for physical substrate dimensions, both in the central part and at the tails of the distribution.
\end{abstract}

\maketitle

\section{Introduction}

The collective properties that characterize dynamical complex systems often emerge from the interplay 
between external driving and dissipation, in such a way that frequently criticality holds spontaneously in large regions of parameter space \cite{Grinstein1991,Odor2004,Pruessner12,Tauber14}. An example is surface kinetic roughening where, in principle, the physical object of interest is a rough surface or interface whose time evolution results from the interplay between deterministic and stochastic processes as for, e.g., thin film or epitaxial growth, or for bacterial colonies \cite{Barabasi95,Krug97}. Notably, the critical behavior displayed by this type of systems is proving to be ubiquitous across system nature and physical scales, to the extent that some of its main instances ---like the celebrated one-dimensional (1D) Kardar-Parisi-Zhang (KPZ) universality class \cite{KPZ86,Kriecherbauer10,Halpin15,Takeuchi18}--- are recently becoming relevant even to non-interfacial systems. Recent examples can be found from quantum \cite{Fontaine2022,Wei2022,Sieberer2025} to active \cite{Caballero2018b,Caballero2020} matter, the kinetics of chemical reactions \cite{Mondal2022}, or the synchronization of phase or limit cycle \cite{Gutierrez2023,Gutierrez2023b,Gutierrez2024} oscillators. 

Actually, work on surface kinetic roughening, and in particular that related with the KPZ universality class \cite{KPZ86,Halpin15,Takeuchi18}, is recently expanding our understanding of critical phenomena at large. Indeed, the critical behavior that ensues is quite rich, as it implies universal behavior not only from the point of view of the values of the critical exponents, but also with respect to the probability distribution function (PDF) of one-point field fluctuations along the time-dependent regime of evolution \cite{Kriecherbauer10,Halpin15,Takeuchi18}. Thus, for 1D KPZ interfaces, the PDF is a member of the celebrated Tracy-Widom (TW) family of probability distributions \cite{Bertin2006,Makey20}, which generalizes naturally into other PDF families in higher dimensions, as has been experimentally confirmed for turbulent liquid crystals \cite{Takeuchi2010,Takeuchi18} and thin film growth \cite{Almeida14,Halpin-Healy2014,Orrillo17} in the 1D and 2D cases, respectively. For the KPZ class, these PDF are conspicuously non-Gaussian, e.g., their non-zero skewness being related with the occurrence of a preferred growth direction \cite{Krug97} and the irreversible character of the dynamics. 

Analogous behavior on the occurrence of universal PDF in the growth regime has been assessed for other universality classes as exemplified by the conserved KPZ equation \cite{Carrasco2016}, to the extent that, currently, the PDF of field fluctuations is considered as an important trait to unambiguously characterize the kinetic roughening universality class (see, for instance, recent discussions in Refs.\ \cite{Rodriguez21,Marcos22} and others therein). Interestingly, this and other properties of the critical behavior of kinetically rough surfaces (like, for example, novel forms of dynamic scaling ans\"atze) have been uncovered for more standard dynamical critical systems only very recently, as in the non-conserved critical dynamics of the 2D Ising universality class \cite{Vaquero2025}, including experimental verification \cite{Allemand2025}.

The kinetic roughening universality classes just mentioned correspond to systems in which the main source of randomness is time-dependent noise, whose competition with driving and dissipation at comparable time scales induces the so-called generic scale invariance alluded to above \cite{Grinstein1991,Tauber14}. However, in many other contexts noise is quenched and does not evolve in the time scales of evolution for the interface, as in the physical examples of fluid flow or of an elastic manifold evolving in disordered media \cite{Barabasi95}. In these systems, scale invariance also frequently ensues, with eventual connections to so-called self-organized criticality \cite{Odor2004,Pruessner12,Wiese2022}. 
Back to surface kinetic roughening proper, one of the main universality classes in the presence of quenched disorder \cite{Barabasi95,Odor2004,Wiese2022} is that of the quenched KPZ (qKPZ) equation \cite{Galluccio1995,Stepanow1995},
\begin{equation}
    \frac {\partial h(\rr,t)}{\partial t} = F + \nu\nabla^2h(\rr,t) + \frac {\lambda}2[\boldsymbol \nabla h(\rr,t)]^2 + \eta(\rr, h) ,
    \label{eq:qkpz}
\end{equation}
where $h(\rr,t)$ measures the height of an interface above position $\rr \in \mathbb{R}^d$ on a $d$-dimensional substrate at time $t$, $F$ is a constant external driving force, $\nu > 0$, $\lambda$ are additional parameters, and $\eta(\rr, h)$ is uncorrelated, zero-average, Gaussian quenched disorder with amplitude $D > 0$, such that
\begin{equation}
    \langle \eta(\rr, h)\rangle =0\,,
    \label{eq:qkpznoide1}
\end{equation}
\begin{equation}
    \langle \eta(\rr_1, h_1) \eta(\rr_2, h_2)\rangle =2 D \delta^d(\rr_1-\rr_2) \delta(h_1-h_2)\,,
    \label{eq:qkpznoide2}
\end{equation}
where $\delta^d(\boldsymbol{u})$ denotes the Dirac delta function in dimension $d$. 

This model has a very rich dynamical behavior that has been fully elucidated recently in experiments of domain walls in magnetic films \cite{Moon2013} and of reaction fronts in disordered media \cite{Atis2015}. Here, we will concern ourselves with the $\lambda>0$ case only,\footnote{For $\lambda <0$, Eq.\ \eqref{eq:qkpz} also describes a pinning transition, but for a very different class (so-called, negative qKPZ) of faceted interfaces \cite{Jeong1996,Jeong1999,Atis2015}.} where two different universality classes are identified. Thus, a pinning-depinning transition exists at a non-zero critical value of the driving force $F=F_c$, such that the interface is pinned, i.e.,\ the average velocity is zero, or moving (non-zero average velocity) for $F\leq F_c$ or $F>F_c$, respectively \cite{Barabasi95,Leschhorn1996,Ramasco01}. The front described by Eq.\ \eqref{eq:qkpz} only displays the exponent values that define the qKPZ universality class exactly at depinning when $F=F_c$ \cite{Leschhorn1996}. For $F\gg F_c$ and $\lambda>0$, the scaling exponents are those of the (standard) KPZ universality class with time-dependent noise \cite{Tang1995,Leschhorn1996}. For comparison, the scaling behavior of the latter KPZ equation [obtained by replacing the disorder $\eta(\boldsymbol{r},h)$ in Eq.\ \eqref{eq:qkpz} by similarly uncorrelated, Gaussian, time-dependent noise $\eta(\boldsymbol{r},t)$] depends neither on the value of $F$ nor on the sign of $\lambda$ \cite{Barabasi95,Krug97}.

Part of the interest generated by the qKPZ equation is due to its relation to directed percolation (DP), another paradigm of non-equilibrium phenomena \cite{Henkel2008}. This connection was much developed early on through work on the discrete directed percolation depinning (DPD) model, believed to be in the qKPZ universality class and initially formulated to describe experiments on front propagation caused by fluid imbibition in paper \cite{Amaral95,Barabasi95}. 

A mapping between the DPD model and DP allowed to interpret the critical exponents of DPD in terms of those of DP, in principle for any value of $d$. Much work has been devoted since to assessing the rich set of critical exponents and behaviors (like the statistics of avalanches \cite{Chen2011}, crossover behavior \cite{Chen2015} and others) for this and other discrete models in the qKPZ universality class, and for the equation itself, see Ref.\ \cite{Wiese2022} for a review. Very recently, further numerical and analytical works \cite{Mukerjee2023,Mukerjee2023b} have revised previous arguments and exponent values in the literature \cite{Amaral95}, concluding in particular that the dynamic exponent $z$ does not derive directly from the assumed identification between the length of the invading path with the duration of an avalanche and reporting a different value for the exponent.

In this paper, we revisit the kinetic roughening properties of fronts in the qKPZ universality class at depinning in the physical substrate dimensions $d=1$ and 2. In parallel with recent developments for the case of time-dependent noise, we aim to assess the existence of analogous universal statistics of the front fluctuations in the growth regime, namely, a unique PDF valid for all times prior to saturation to steady state, provided that fluctuations are suitably rescaled to compensate for their time-increasing amplitude. In this process, we compute directly the full set of critical exponents characterizing the kinetic roughening behavior without resorting to mappings and/or scaling relations and compare with well-known values as well as with more recent results.

The paper is organized as follows. After this introduction, Sec.\ \ref{sec:model} contains the description of the model we study, the various quantities that will be employed for its study, and some simulation details. Our numerical results for the critical exponent values in $d=1$ and $d=2$ are reported in Secs.\ \ref{sec:res1d} and \ref{sec:res2d}, respectively. This is followed by our results for the universal PDF in $d=1, 2$, described in Sec.\ \ref{sec:fluct}. Section \ref{sec:disc} provides a discussion of our results in the perspective of previous literature, while Section \ref{sec:concl} contains a summary and the conclusions of our work. Finally, a number of additional numerical results are collected into two appendixes at the end.

\section{Model and Observables}
\label{sec:model}

\subsection{The discrete model}
The calculations reported herein have been performed on a lattice of length $L$ or on a square lattice of dimensions $L \times L$ for the $d = 1$ and $d = 2$ cases, respectively, subject to periodic boundary conditions (PBC). Following the discrete model (DM) put forward in Ref.\ \cite{Song07}, the local height $h(\rr,t)$ at a given time $t$ evolves according to
\begin{equation}
    h(\rr, t+1) = \begin{cases} \begin{matrix} h(\rr,t) + 1 & \text{ for }\;  g(\rr,t)>0,\\ h(\rr, t) & \text{ for } \; g(\rr, t) \leq 0, \end{matrix} \end{cases}
    \label{eq:def_automaton}
\end{equation}
where
\begin{equation}
    g(\rr,t)= F + \nabla^2h(\rr,t)+\frac 12[\boldsymbol{\nabla} h(\rr,t)]^2 + \eta(\rr,h),
    \label{eq:g}
\end{equation}
corresponding to the qKPZ equation, Eq.\ \eqref{eq:qkpz}, with $\nu = \lambda = 1$ for simplicity. 

In Eq.\ \eqref{eq:g}, the noise has been uniformly chosen in the proper ranges to ensure that the system is in the strongly non-linear regime. Specifically, we have drawn the random numbers uniformly within the $(-5, 5)$ and $(-15, 15)$ intervals for $d = 1$ and $d = 2$, respectively. 

In addition, the Laplacian term has been discretized as
\begin{equation}    
    \nabla^2 h(\rr,t)=\sum_{\boldsymbol{s} \in \text{NN}(\boldsymbol{r})}\left[h(\boldsymbol{s},t)-h(\rr,t)\right],
    \label{eq:discrete_laplacian}
\end{equation}
where the sum spans the nearest neighbors of site $\boldsymbol{r}$, denoted as $\text{NN}(\boldsymbol{r})$. Finally, a centered-difference approach for the gradient has been taken, as in Refs.\ \cite{Lee05,Song07,Mukerjee2023}. For the 1D model ($d=1$), we write
\begin{equation}
    \left[\boldsymbol{\nabla} h(x,t)\right]^2 = \left(\frac {h(x-1,t)-h(x+1,t)}2 \right)^2
    \label{eq:gradient_1d}
\end{equation}
and, for the 2D model ($d=2$),
\begin{align}
  \nonumber
    \left[\boldsymbol \nabla h(\rr, t)\right]^2 &= \left(\frac {h(x+1,y,t)-h(x-1,y,t)}2\right)^2 \\
    &+ \left(\frac {h(x,y+1,t)-h(x,y-1,t)}2\right)^2\,.
\end{align}

We have used a flat surface [i.e., $h(\rr,t=0)\equiv 0$] as initial condition for all the realizations of the noise. 

\subsection{Observables}
\label{sec:obs}
Several observables have been defined to describe the dynamics of the aforementioned model; most of these observables are standard and have been previously used in many studies. The average front position is defined as
\begin{equation}
    \overline{h}(t) \equiv \langle \overline{h}(t)\rangle \equiv \left\langle \frac 1{L^d} \sum_{\rr}h(\rr,t)\right\rangle,
    \label{eq:def_front}       
\end{equation}
where $V=L^d$ is the volume of the $d$-dimensional lattice and the sum spans all its sites. In Eq.\ \eqref{eq:def_front}, the $\overline{(\cdots)} \equiv (1/L^d)\displaystyle\sum_{\textbf{r}} (\cdots)$ notation accounts for the space average over the lattice and $\langle (\cdots) \rangle$ for the average over different realizations of the noise. From Eq.\ \eqref{eq:def_front}, the average front velocity may be straightforwardly defined as the time derivative of the average front position, i.e., 
\begin{equation}
v(t) = \frac {d\overline{h}(t)}{dt}\,.
\end{equation}
Near the pinning transition ($F\gtrsim F_c$), the asymptotic front velocity varies with the applied force as 
\begin{equation}
    v(F) \sim (F - F_c)^\theta,
    \label{eq:v_F}
\end{equation}
where $F_c$ is the critical force, $\theta$ is the so-called velocity exponent, and we have neglected subdominant terms (scaling corrections). For $F = F_c$, on the other hand, the front velocity is a decreasing function of time, with the dominant term being given by \cite{Lee05,Song07}
\begin{equation}
    v(F=F_c,t) \sim t^{-\delta},
    \label{eq:v_vs_t}
\end{equation}
where $\delta$ is another critical exponent. 

Our next relevant observable is the front width (or roughness), which is defined as the standard deviation of the front position around its average, namely,
\begin{equation}
    w^2(t) = \left\langle \overline{[h(\rr, t) - \overline h(t)]^2}\right\rangle \,.
    \label{eq:def_width}
\end{equation}

Under kinetic roughening conditions, the roughness follows the so-called Family-Vicsek (FV) dynamic scaling law 
\begin{equation}
    w(t) = t^\beta f(t/L^z),
    \label{eq:FV}
\end{equation}
where $\beta$ and $z$ are called growth and dynamic exponents, respectively. In Eq.\ \eqref{eq:FV}, the scaling function $f(y) \sim \text{const.}$ for $t \ll L^z$, so that $w(t \ll L^z) \sim t^\beta$, and $f(y) \sim y^{-\beta}$ for $t \gg L^z$, so that $w(t \gg L^z) \sim \text{const.} \equiv w_{\mathrm{sat}}$. The saturation roughness scales with the size of the system, $L$,  as $w_{\mathrm{sat}} \sim L^\alpha$, where $\alpha$ is the roughness exponent, and is reached after a saturation time $t_{\rm sat} \sim L^z$. Note that, as a consistency condition of the FV ansatz for the roughness, the $\alpha$, $\beta$, and $z$ exponents are not independent: Indeed, $L^{\alpha} \sim w_{\rm sat} \sim t_{\rm sat}^{\beta} \sim L^{z \beta}$, hence $\alpha=z\beta$ \cite{Barabasi95}. 

Likewise, for the qKPZ universality class $\delta$ and $\beta$ are related to each other, since $\delta + \beta = 1$ as suggested by an interpretation of $v(t)$ in terms of the density order parameter in phase transitions with absorbing states \cite{Dickman2000}, and verified for qKPZ at depinning \cite{Lee05,Song07}. Indeed, provided that $\bar{h}(t) \sim w(t)$, then $1-\delta=\beta$, since $v \sim \bar{h}/t$ \cite{Dickman2000,Barreales2023}. Note that this scaling relation is {\em not} satisfied, e.g., by the KPZ universality class with time-dependent noise, where $\delta=0$ but $\beta=1/3$, or by other kinetic roughening universality classes that feature non-trivial values of $\delta$. For instance, a continuum model for the spreading of a precursor film in band geometry features $\delta=0.42$ and $\beta=0.35$ \cite{Marcos22}.

As for the direct physical meaning of the kinetic roughening exponents, $\beta$ rules the growth with increasing time of the fluctuations of the front around the average front position, as evidenced by Eq.\ \eqref{eq:def_width}, while $\alpha$ is related to the fractal dimension of the front \cite{Barabasi95,Mozo22}. Finally, the dynamic exponent $z$ is related to the time increase of the lateral correlation length $\xi(t)$ along the front, i.e.,
\begin{equation}
    \xi(t) \sim t^{1/z} \,.
    \label{eq:def_xi}
\end{equation}

To account for statistical two-point correlations within the system, an additional function is useful, namely the height-difference correlation function $C_2(\rr, t)$ defined as
\begin{equation}
    \begin{split}
        C_2(\rr,t) = \frac{1}{L^d} \sum_{\boldsymbol{x}} \left\langle[h(\boldsymbol{x+r},t) - h(\boldsymbol{x},t)]^2\right\rangle \\ =2\langle\overline {h(t)^2}\rangle - \frac 2{L^d}\sum_{\boldsymbol{x}}\langle h(\boldsymbol{r+x},t)h(\boldsymbol{x},t)\rangle\,.
    \label{eq:def_C2}
\end{split}
\end{equation}
Under standard kinetic roughening conditions where the FV dynamic scaling applies, this function satisfies the relation
\begin{equation}
    C_2(\rr, t) = r^{2\alpha}g\big(r/\xi(t)\big),
    \label{eq:C2_scaling}
\end{equation}
where $r=\|\rr\|$ and the scaling function $g(u) \sim u^{-2\alpha}$ for $u \gg 1$, while $g(u) \sim \text{const.}$ for $u \ll 1$. Note that the latter behavior implies that, as intuitively expected, $C_2(\rr,t)$ becomes independent of $r$ for distances larger than the correlation length value at the given time, i.e., for $r \gg \xi(t)$. 

The height-difference correlation function allows one to compute the correlation length $\xi(t)$ explicitly, as has been reported elsewhere \cite{Barreales20}. Essentially, the correlation length can be thought of as the distance along the substrate coordinate at which the $C_2(\rr,t)$ function takes on a significant fraction of its value at the plateau value $C_2(L/2,t)$ (recall that we are considering PBC). Thus, 
\begin{equation}
    C_2(\xi_a(t), t) = a  C_2(L/2,t),
    \label{eq:a_scaling}
\end{equation}
where $a = 0.9$ in this work; the scaling behavior of the correlation length is not much affected by the particular value of $a$ \cite{Barreales20}.

Finally, we have computed the probability density function (PDF) of the front fluctuations in the growth or time-dependent regime, i.e., for times $t \ll L^z$ so that $w \sim t^\beta$. Specifically, front fluctuations $\chi(\rr,t)$ are defined as
\begin{equation}
\chi(\rr,t)=\frac{h(\rr,t)-\overline h(t)}{w(t)} .
\label{eq:def_fluctuations}
\end{equation}
Thus, fluctuations are centered around the average front and their time-dependent amplitude [as estimated by the roughness $w(t)$] is crucially normalized out. In this way, the ensuing PDF becomes stationary (i.e., the form of the PDF becomes time-independent) and universal in the sense that the same stationary PDF occurs for all members of the same kinetic roughening universality class, as has been verified in kinetic roughening systems \cite{Kriecherbauer10,Halpin15,Takeuchi18,Carrasco2016,Barreales20,Rodriguez21,Marcos22,Barreales2023} and, recently, for more standard dynamical critical systems as well \cite{Vaquero2025}. As noted above, to date the growth PDF is considered an important trait to unambiguously identify the kinetic roughening universality class, beyond the particular values of the scaling exponents.

The errors of all quantities reported in this work have been computed following the jackknife procedure, especially suitable for highly correlated data (e.g., all the fits performed in time) \cite{Efron82,Young15}; see Refs \cite{Michael94, yllanes:11} and Appendix B of Ref.~\cite{Barreales20} for additional details. The system sizes, maximum times achieved, and number of realizations of the noise (or number of runs) considered for both models are collected in Table~\ref{tab:details} of Appendix~\ref{appen:details}.

\section{1D critical exponents}
\label{sec:res1d}

In this section, we report our results for the 1D model regarding the critical force and its related exponent $\theta$, the growth exponents $\alpha$, $\beta$, and $\delta$, and the behavior of the height-difference correlation function $C_2(r,t)$, which will allow us to compute the correlation length $\xi(t)$ and its associated exponent $z$.

\subsection{Evaluation of the critical force}
\label{sec:res1d:Fc}

\begin{figure}[!t]
    \centering \includegraphics[width=\columnwidth]{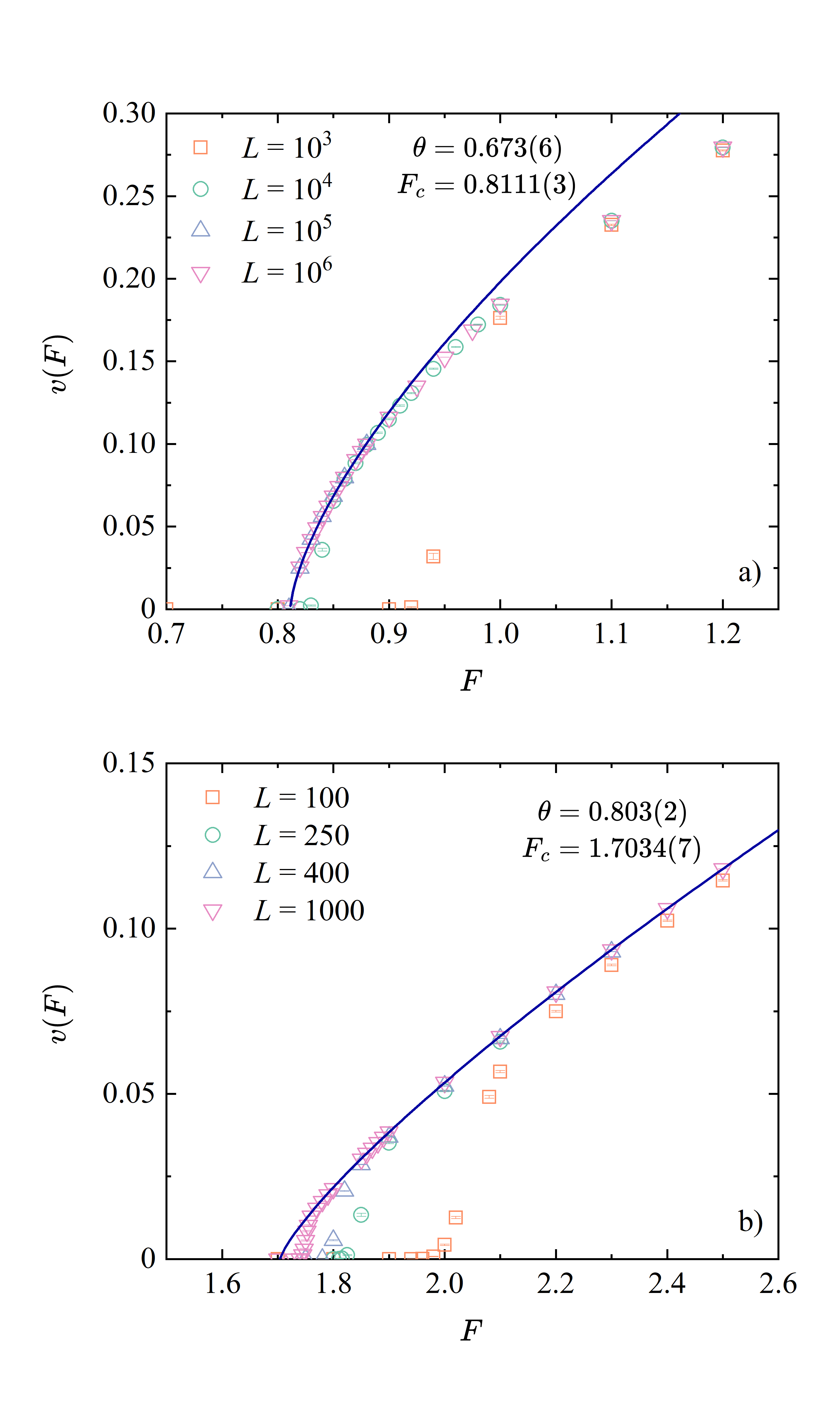}
    \caption{\textit Steady-state front velocity vs.\ $F$ for different values of $L$ as indicated in the legends, for (a) $d = 1$ and (b) $d=2$. The solid lines correspond to $F\sim(F-F_c)^{\theta}$ for the values of $F_c$ and $\theta$ indicated in the corresponding panel.}
    \label{fig:vF}
\end{figure}

The approach adopted in this work consists of evaluating the front
velocity for various values of the external driving force
$F$. According to Eq.~\eqref{eq:v_F}, the fit of these results to a
power law yields an estimation of both the critical force
$F_c$ and the velocity exponent. Thus, Fig.~\ref{fig:vF}
plots the velocity values vs.\ $F$, for different system
sizes, computed for $d = 1$ and $d = 2$. In both cases, one observes that the front remains pinned (i.e., with zero velocity) for forces below some threshold which is dependent on the system size, specifically decreasing as $L$ increases. 

For the 1D model [see Fig.~\ref{fig:vF}(a)], we find $F_c=0.8111(3)$ and $\theta=0.673(6)$ for the largest size considered in this analysis $L=10^6$,
\footnote{For our largest value of  the lattice size ($L_\text{max}=10^6$) we determine what values of the velocity are independent of the size of the system (i.e., $v$ is asymptotic, independent of $t$ and $L$): this happens for $F\ge F_\text{min}$, with $F_\text{min}$ depending on $L_\text{max}$. We perform fits to the scaling law, Eq.\ \eqref{eq:v_F}, in the interval $[F_\text{min}, F]$ starting from the higher simulated value of $F$ and reducing it until we obtain a $p$-value~\cite{Young15} larger than 5\%. We then continue the analysis by further increasing the minimum force in order to verify the stability of the exponent $\theta$ and the critical force $F_c$.}
which we take as representative of the asymptotic limit.\footnote{For uncorrelated data, all the fits reported in this paper have a $p$-value larger than 5\%.} Our values of $\theta$ and $F_c$ compare very well with those of Ref.\ \cite{Song07}, where $F_c=0.8105(5)$ and $\theta\approx 0.672$ were reported for the same model and disorder strength. Furthermore, our estimate of $\theta$ is consistent with the 1D DPD model, for which $\theta = 0.66(4)$ \cite{Amaral95}. Notice that, in principle, the value of $F_c$ is not expected to be universal, since it depends on the disorder strength implemented in the automaton model.

In the remainder of this paper, all data reported, for both the $d = 1$ and $d = 2$ cases, were obtained for an external force equal to the corresponding critical force for the pinning-depinning transition, i.e., $F = F_c$. 

\subsection{Growth exponents}
\label{sec:res1d:exponents}

\begin{figure}[!t]
    \centering    
    \includegraphics[width=\columnwidth]{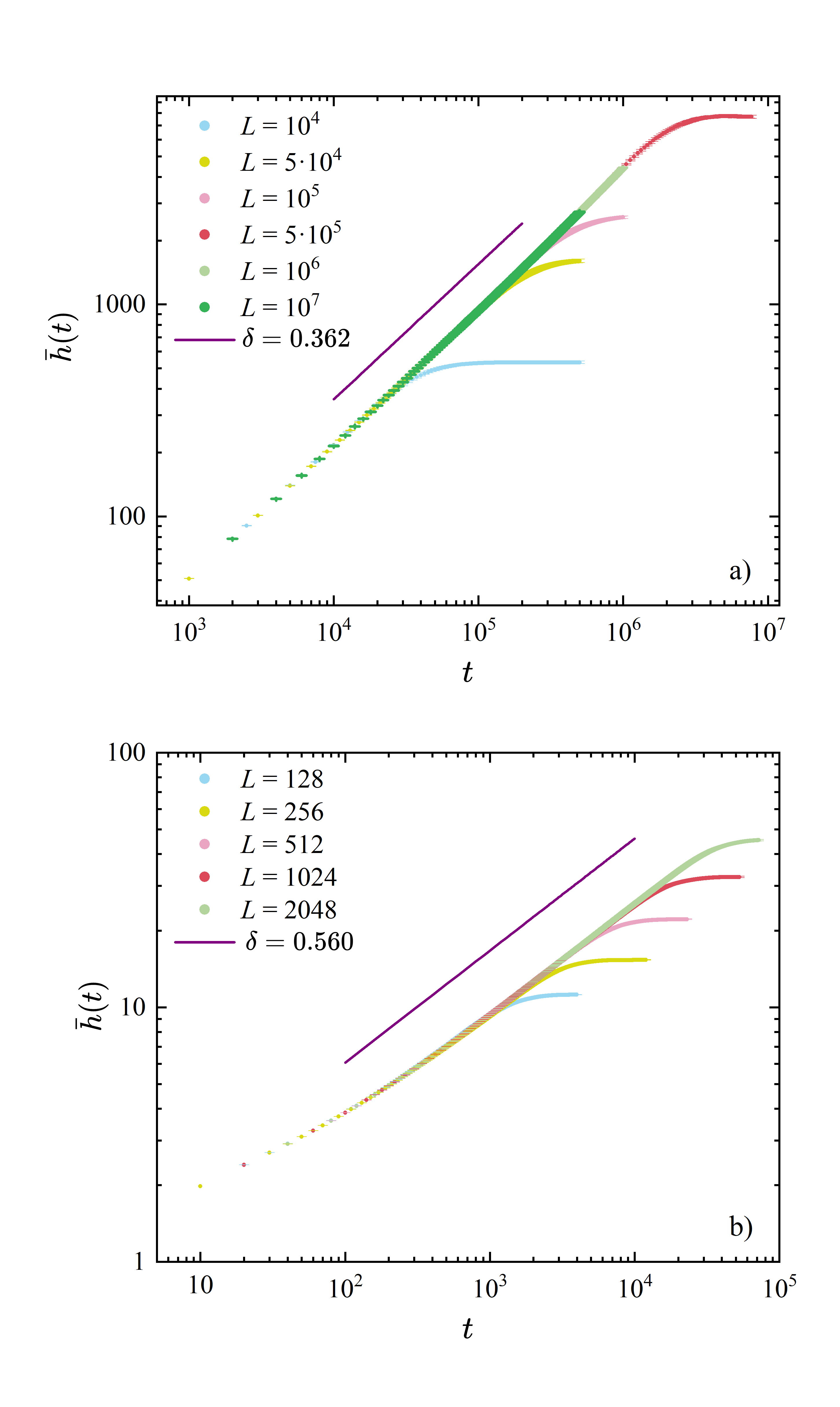}
    \caption{Average front vs.\ time for different values of $L$, as indicated in the legends, for (a) $d = 1$ and (b) $d=2$. The straight solid lines correspond to $\overline{h}(t)\sim t^{1-\delta}$ computed using the value of $\delta$ obtained for the largest system size in each case, see legends and Table \ref{tab:exponents1D}.} 
    \label{fig:h_vs_t}
\end{figure}

Taking into account Eq.~\eqref{eq:v_vs_t}, the average position of the front at the critical force is expected to evolve with time as $\bar{h}(t)\sim t^{1-\delta}$. Figure \ref{fig:h_vs_t}(a) shows $\bar{h}(t)$ as a function of time for several values of the system size and $d = 1$; the corresponding values for the $\delta$ exponent are reported in Table \ref{tab:exponents1D}. Using our largest lattice, we find $\delta=0.362(1)$ for $d=1$; note that the analysis of $L\ge 5 \times 10^5$ runs yields the same estimate.

\begin{table*}[!t]
\begin{ruledtabular}
\begin{tabular}{c | c  c  c c | c  c  }

$L$ & $\delta$ & $\beta$ & $z$ & $\alpha$ &$\beta+\delta$ & $\beta z$  \\

\hline
$10^4$ & 0.364(5) & 0.480(7) & 1.036(11) & 0.638(15) & 0.844(9)  & 0.50(2) \\ 
$5\times 10^4$ & 0.360(4) & 0.528(6) & 1.02(3) & 0.630(19) &0.888(8) &  0.54(2)\\
$10^5$ & 0.371(6) & 0.580(3) & 1.00(2) & 0.626(9)& 0.951(7) & 0.58(2)\\
$5\times 10^5$ & 0.363(2) & 0.613(2) & 1.00(2) & 0.644(7) & 0.976(3) & 0.607(12)\\
$10^6$ & 0.363(1) & 0.623(1) & 1.016(7) &  0.640(11)  & 0.986(1) &  0.633(4) \\
$10^7$ & 0.362(1) & 0.6290(6) & 1.017(13)$^*$& 0.642(7)$^*$& 0.991(1) & 0.640(8)$^*$  \\
\end{tabular}
\end{ruledtabular}
\caption{Critical exponent values for the 1D model. We stress that the $\delta$, $\beta$, $z$, and $\alpha$ exponents have been computed directly (i.e., without using scaling relations). Besides these four exponents, we checked the scaling relations $\beta+\delta=1$ and $\beta z=\alpha$ (last two columns of the table). We recall that the value of $\alpha$ obtained using the scaling of the plateaus of the roughness is $\alpha=0.6325(17)$. We have been able to simulate, as a control,  a small number of initial conditions (100) for $L=10^7$ in order to compute the $z$ and $\alpha$ exponents. The results marked with an asterisk are compatible with those for $L=10^6$, but have a higher error for $z$. Therefore, our final estimate for $z$ is that computed for $L=10^6$.}
\label{tab:exponents1D}
\end{table*}

\begin{figure}[!t]
    \centering    
    \includegraphics[width=\columnwidth]{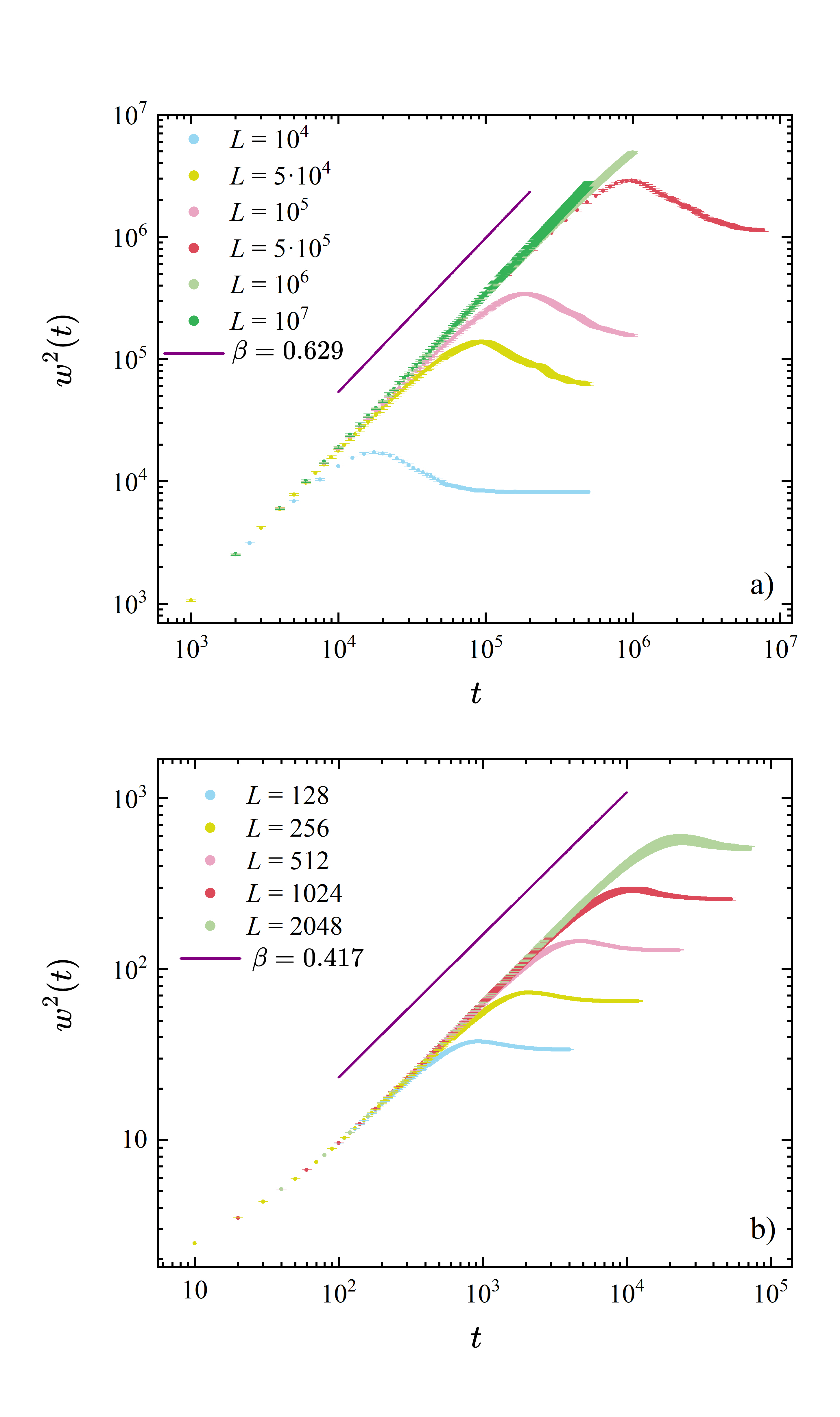}
    \caption{Squared roughness vs.\ time for different values of $L$, as indicated in the legends, for (a) $d = 1$ and (b) $d=2$. The straight solid lines correspond to $w(t)\sim t^{\beta}$ using the value of $\beta$ obtained for the largest system size in each case, see legends and Table \ref{tab:exponents1D}.} 
    \label{fig:w_vs_t}
\end{figure}

On the other hand, Fig.~\ref{fig:w_vs_t}(a) displays the squared roughness data versus simulation time, as computed for different system sizes and $d = 1$. In all cases a growth regime characterized by a well-defined value of $\beta$ follows an initial transient. The growth regime ends at maxima of the $w^2(t)$ curves, from where the squared width decreases to a time-independent saturation value. The fit of the numerical $w(t)$ data to Eq.\ \eqref{eq:FV}, in the growth time interval where the condition $t \ll L^z$ is satisfied, yields the $\beta$ exponent values recorded in the third column of Table~\ref{tab:exponents1D} for $d = 1$. This exponent increases slightly with $L$, reaching a value $\beta = 0.6290(6)$ for
$L=10^7$. In relation to this, the sixth column of Table~\ref{tab:exponents1D} shows that the sum $\beta+\delta$ is approaching one as $L$ increases. In particular, for $L = 10^7$, we obtain $\beta+\delta=0.991(1)$ which is in good agreement with the theoretical expectation \cite{Lee05,Song07}.

\begin{figure}[!t]
    \centering    
    \includegraphics[width=\columnwidth]{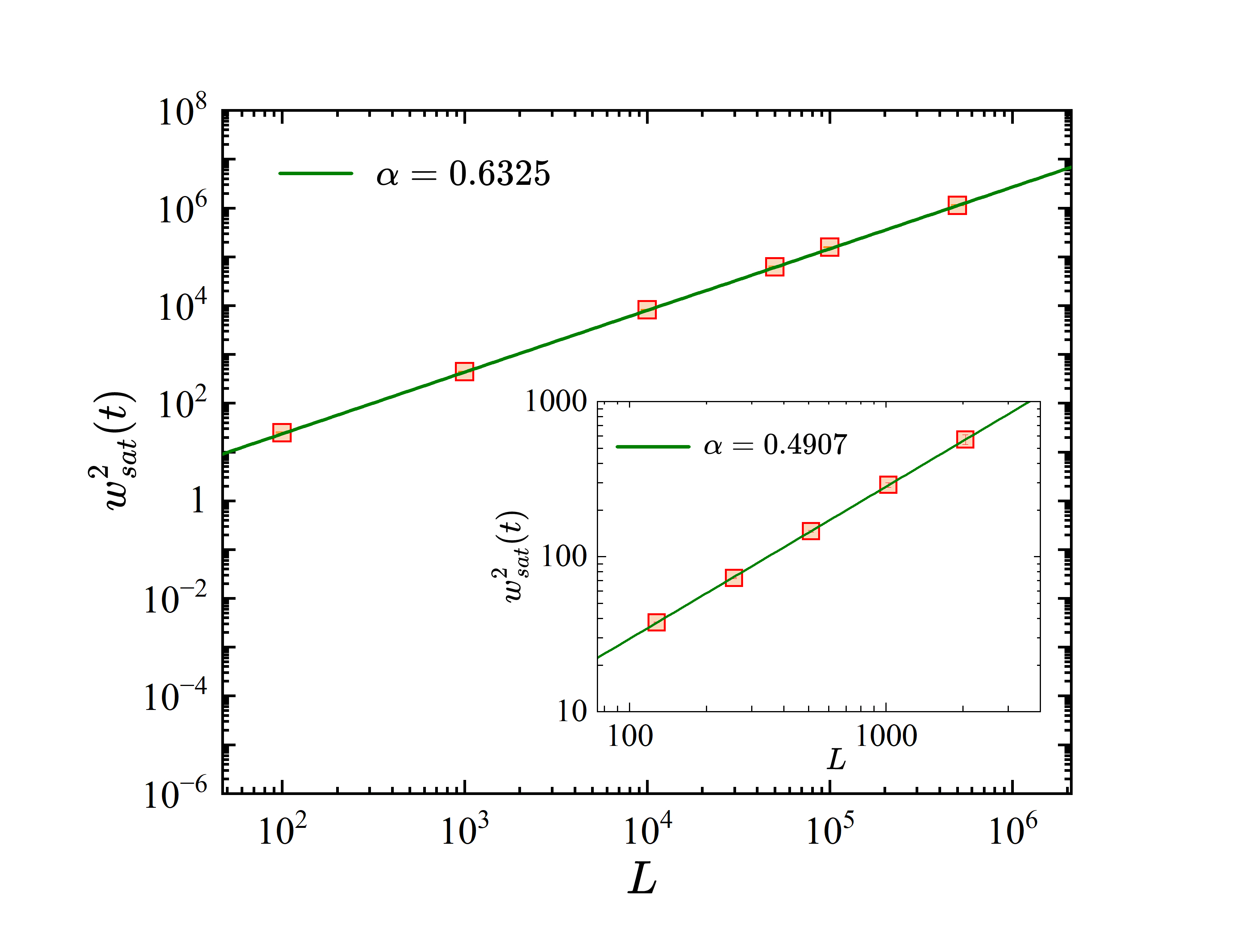}
    \caption{Logarithmic plot of the saturation squared roughness $w^2_\text{sat}$ vs.\ $L$ for the 1D (main panel) and 2D (inset) models. According to Eq.~\eqref{eq:FV}, the slope of the best-fit straight line in the plot equals 2$\alpha$, with $\alpha$ as indicated in each legend.}
    \label{fig:wsat_vs_L}
\end{figure}

The roughness exponent may also be estimated from the data in Fig.~\ref{fig:w_vs_t}(a), since $w_\text{sat} \sim L^\alpha$, obtaining $\alpha=0.6325(17)$, see Fig.\ \ref{fig:wsat_vs_L}.\footnote{Notice that the data used in this fit are fully uncorrelated since they have been originated in runs with different lattice sizes.}

\subsection{Correlation function}
\label{sec:res1d:C2}

As mentioned above, the study of the height-difference correlation function $C_2(\rr,t)$ provides significant local information on
the dynamic behavior of the growing front. In Appendix \ref{appen:scaling} we show $C_2(\rr,t)$ as a function of $r$ for several values of time and sizes, both for $d = 1$ and $d = 2$. The qualitative behavior of the curves fully conform with Eq.\ \eqref{eq:C2_scaling} and FV dynamic scaling behavior.

\begin{figure}[!ht]
    \centering
    \includegraphics[width=\columnwidth]{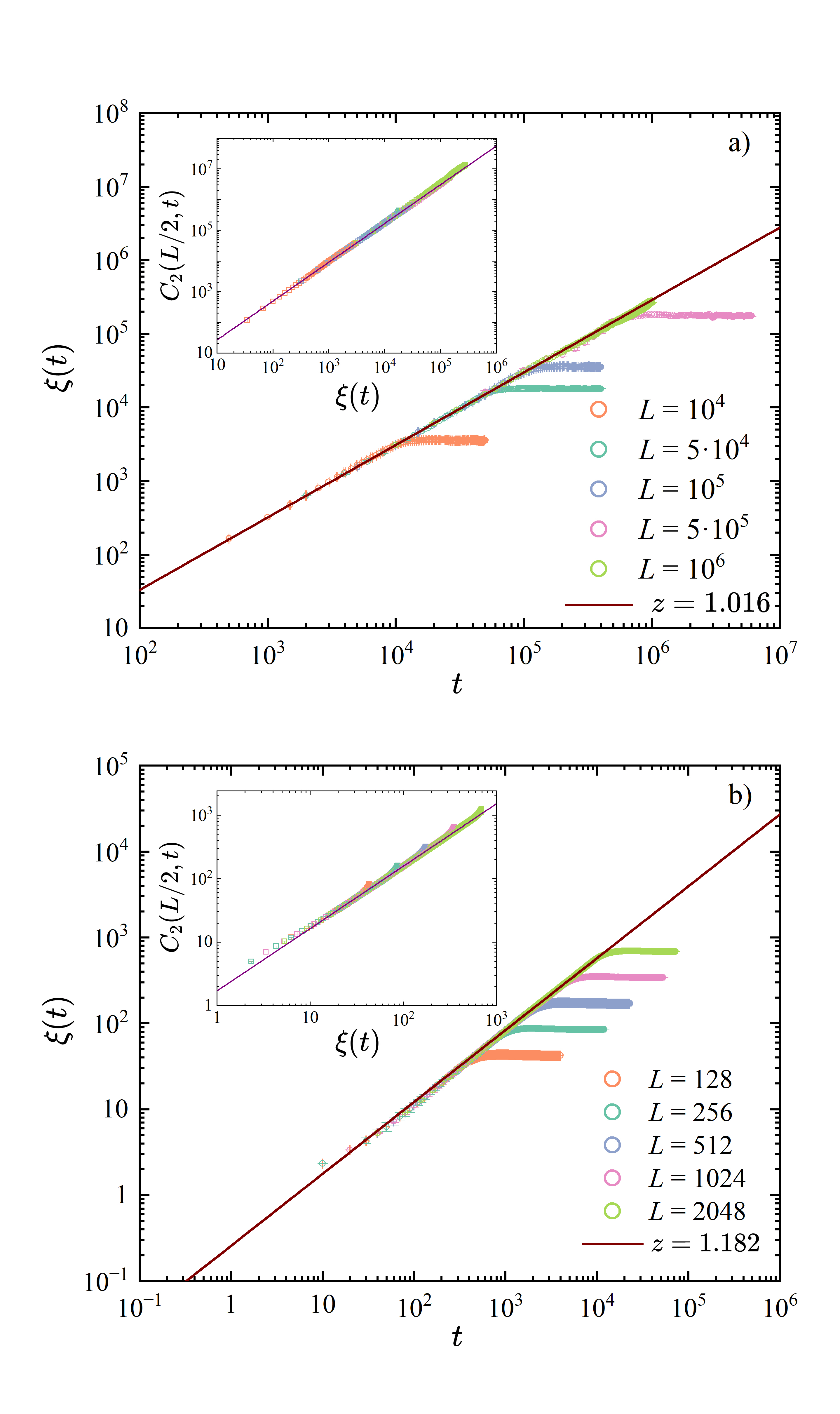}
    \caption{Correlation length computed from Eq.~\eqref{eq:a_scaling} with $a = 0.9$, for different system sizes $L$ as indicated in the legends, (a) $d = 1$ and (b) $d = 2$. The straight lines correspond to $\xi(t) \sim t^{1/z}$ with $z$ as indicated in the legend, which corresponds to the largest system size in each case, see Table \ref{tab:exponents1D}. In both figures, the insets plot the $C_2(L/2,t)$ function vs.\ $\xi(t)$; these plots yield straight lines with slope 2$\alpha$, see the text.}
    \label{fig:chi_vs_t}
\end{figure}

Let us first consider the correlation length, $\xi(t)$, which, as discussed above, can be regarded as the distance
at which the height-difference correlation function takes on the 90\%
of its value at the plateau [\textit{cf.} Eq.~\eqref{eq:a_scaling}]. 
Figure \ref{fig:chi_vs_t} shows the thus computed
$\xi(t)$ values for the dimensionalities and system sizes considered in this
work. In all cases, the same trend is observed, analogous to that for the
front: an initial transient is followed by a well-defined growth regime that ends in the saturation of the correlation length when $\xi(t)$ reaches a value similar to the size of the lattice. This trend is fully
consistent with the expected behavior, Eq.\ \eqref{eq:a_scaling},
and allows one to compute the $z$ exponent. As is the case with the exponent $\beta$, the value of the dynamic exponent $z$ exhibits a slight dependence on $L$, see Table~\ref{tab:exponents1D}. Our final estimate is $z=1.016(7)$ for $L=10^6$.

The data from Fig.~\ref{fig:chi_vs_t} may be used for an alternative computation of the roughness exponent $\alpha$ based on the behavior of the correlation function at distances larger than $\xi(t)$, see the insets in that figure. Indeed, Eq.~\eqref{eq:C2_scaling} implies that $C_2(\rr,t)\sim \xi^{2\alpha}(t)$ for $r \gg \xi(t)$. Taking into account times such that $\xi(t) \ll L/2$, a logarithmic plot of $C_2(L/2,t)$ as a function of $\xi(t)$ should yield a straight line of slope 2$\alpha$ within some time interval.  In this way, we obtain an estimate of $\alpha$ for each value of $L$ obtaining $\alpha\simeq 0.630$, see Table \ref{tab:exponents1D}, to be compared with our previous estimate (using a much simpler uncorrelated fit which provides a most accurate result) from the scaling of the plateau of the roughness, $\alpha=0.6325(17)$.

\begin{figure}[!h]
    \centering
    \includegraphics[width=\columnwidth]{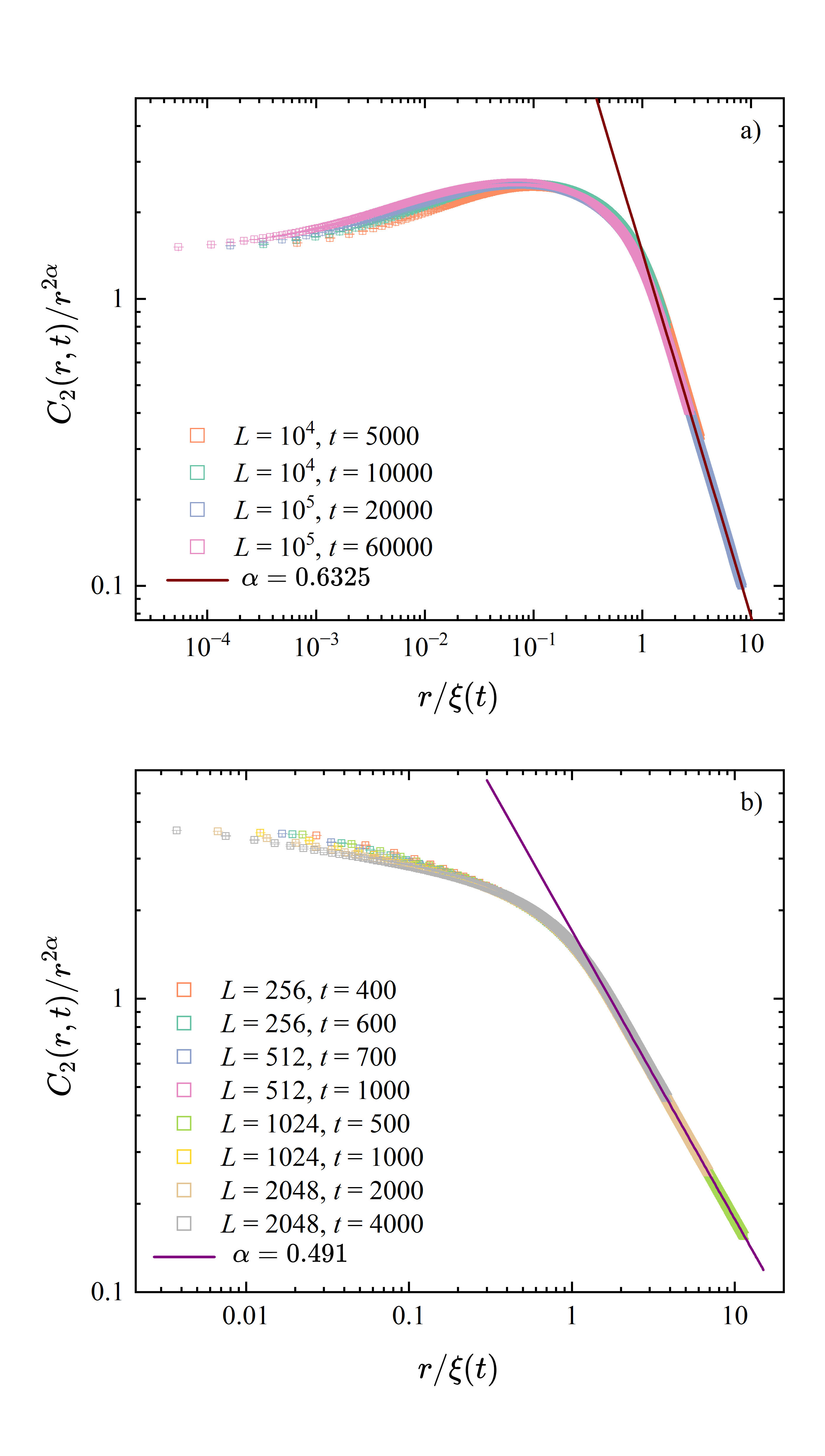}
    \caption{$C_2(\rr,t)/r^{2\alpha}$ vs.\ $r/\xi(t)$ for different values of time and system size $L$ as given in the legends, for (a) $d = 1$ and (b) $d = 2$. The straight lines correspond to the large-argument behavior of the scaling function $g(u)\sim u^{-2\alpha}$, see Eq.\ \eqref{eq:C2_scaling} and the main text, with $\alpha$ as indicated in each legend.}
    \label{fig:C2_collapsed}
\end{figure}

Another check of the goodness of our results also comes from Eq.~\eqref{eq:C2_scaling}, which can be rewritten as
$C_2(\rr,t)r^{-2\alpha} = g[\rr/\xi(t)]$. Consequently, a plot of $C_2(\rr,t)r^{-2\alpha}$ vs.\ $r/\xi(t)$ should yield the universal
(i.e., time-independent) scaling function $g(u)$ with asymptotic behaviors as indicated in Eq.~\eqref{eq:C2_scaling}. Figure \ref{fig:C2_collapsed}(a) shows such a data collapse for $d = 1$, for the selected times and different lattice sizes, in a doubly-logarithmic plot, in good agreement with the theoretical expectations. Note further that the $C_2(\rr,t)$ curves for different times share the same slope at small arguments; as a consequence, there is a range of $r$ where these curves partially overlap for different times. This indicates that the same scaling ansatz holds throughout the lattice, irrespective of the value of $r/\xi(t)$ and, in turn, that a single roughness exponent $\alpha$ suffices to describe the spatial properties of correlations in the system, as indeed expected within the FV dynamic scaling ansatz \cite{Barabasi95, Krug97, Barreales20} (see also Fig.~\ref{fig:C2_unscaled}(a) in Appendix \ref{appen:scaling}).

Finally, we have tested the relation $\alpha=\beta z$ in Table \ref{tab:exponents1D} and we can observe a trend compatible with a final value given by $\alpha=0.6325(17)$. Notice that the $L=10^6$ value is at 1.3 standard deviations of our final estimate.

\section{2D critical exponents}
\label{sec:res2d}

Following the same methodology used in Sec.\ \ref{sec:res1d:Fc}, for our simulations in $d=2$ we obtain $\theta = 0.803(1)$ and $F_c = 1.703(1)$ after fitting the $L=1024$ data to Eq.~\eqref{eq:v_F}, see Fig.~\ref{fig:vF}(b). To our knowledge, this result is new, as we are not aware of previous reports in the literature of the value of $F_c$ in $d = 2$ for the model we are presently studying.

On the other hand, Table~\ref{tab:exponents2D} records our estimates for the remaining scaling exponents ($\delta$, $\beta$, $z$, and $\alpha$) as obtained for several system sizes in $d = 2$, computed using the same methodology as in the 1D model. As expected below the upper critical dimension, the scaling exponent values depend substantially on the dimensionality of the system. In particular, the growth of the average front is now ruled by $\delta = 0.560(3)$, see Fig.\ \ref{fig:h_vs_t}(b), whereas the time behavior of the roughness is described by $\beta = 0.417(2)$ for the largest system size considered in this work, see Fig.\ \ref{fig:w_vs_t}(b). As for $d = 1$, a relatively extended growth regime is observed before pinning, as evidenced in Fig.~\ref{fig:w_vs_t}(b). Note also that the theoretical prediction $\delta + \beta = 1$ is not satisfied for $d = 2$ so accurately as in the 1D case. 

\begin{table*}[t]
\begin{ruledtabular}
\begin{tabular}{c | c  c  c c | c  c }

$L$ & $\delta$ & $\beta$ &  $z$ & $\alpha$ & $\delta + \beta$ & $\beta z$ \\

\hline
128 & 0.587 (6) & 0.382(3) &  1.242(8) & 0.473(4) & 0.969(7) & 0.474(5)  \\
256 & 0.591(4) & 0.382(3) &  1.196(6) & 0.478(3) & 0.973(5) & 0.457(4)  \\
512 & 0.568(6) & 0.406(3) &  1.189(8) & 0.495(3) & 0.974(7) & 0.483(5)  \\
1024 & 0.563(4) & 0.4202(14) &  1.190(7) & 0.499(3) & 0.983(4) & 0.500(3) \\
2048 & 0.560(3) & 0.417(2) &  1.182(4) & 0.491(2) & 0.977(4) & 0.493(3)\\

\end{tabular}
\end{ruledtabular}
\caption{Critical exponents for the 2D model. We stress that the $\delta$, $\beta$, $\alpha$, and $z$ exponents 
have been computed directly (i.e., without using scaling relations) . Besides these four exponents, we checked the scaling relations $\beta+\delta=1$ and $\beta z=\alpha$ (last two columns of the table). We recall that the value of $\alpha$ obtained using the scaling of the plateaus of the roughness is $\alpha=0.4907(16)$.}
\label{tab:exponents2D}
\end{table*}

The $\alpha$ exponent value increases slightly with $L$ as for $d=1$, reaching a maximum value $\alpha = 0.499(3)$ for $L = 1024$. Note that this value, as well as that computed for $L = 2048$, is consistent with the $\beta$ and $z$ exponents computed independently, as shown in the seventh column of Table~\ref{tab:exponents2D}, and with that estimated from the roughness at saturation. Note further that the values of our estimates of the exponents for $L=1024$ and $L=2048$ are compatible within the statistical errors, see Table \ref{tab:exponents2D}. We would also like remark the huge number of initial conditions simulated for the largest lattice.

For $d = 2$, the correlation function $C_2(\rr,t)$ exhibits the same qualitative behavior as in the 1D model, as anticipated by Fig.~\ref{fig:C2_unscaled}(b) in Appendix \ref{appen:scaling}. The $z$ exponent values computed in this case for the system sizes considered in this work are included in Table~\ref{tab:exponents2D}. 
Here again, $z$ decreases slightly with the system size, to reach $z = 1.182(4)$ for $L = 2048$.   

No evidences of anomalous scaling has been observed, such as, for example, the characteristic anomalous upward shift of the $C_2(\rr,t)$ curves for increasing times \cite{Lopez97,Ramasco00,Cuerno04}. This underscores the consistency of the FV dynamic scaling ansatz for the 2D model as well and contrasts with the non-FV, anomalous scaling observed for other systems with absorbing states, see, e.g., Ref.\ \cite{Dickman2000,Barreales2023} and others therein.

\section{Statistics of fronts fluctuations}\label{sec:fluct}

Besides the scaling exponents, the shape of the PDF for fluctuations of the front, defined according to Eq.~\eqref{eq:def_fluctuations} above, contributes substantially to the unambiguous identification of non-equilibrium universality classes, as has been demonstrated in several contexts \cite{Kriecherbauer10,Halpin15,Takeuchi18,Carrasco2016,Barreales20,Rodriguez21,Marcos22,Barreales2023}. 
In particular, the skewness and excess kurtosis have been recently computed in a system closely related to the one we are addressing, namely, the quenched version of the Edwards-Wilkinson equation \cite{Toivonen2022,Sillanpaa2025}.

\begin{figure}[!h]
    \centering
    \includegraphics[width=\columnwidth]{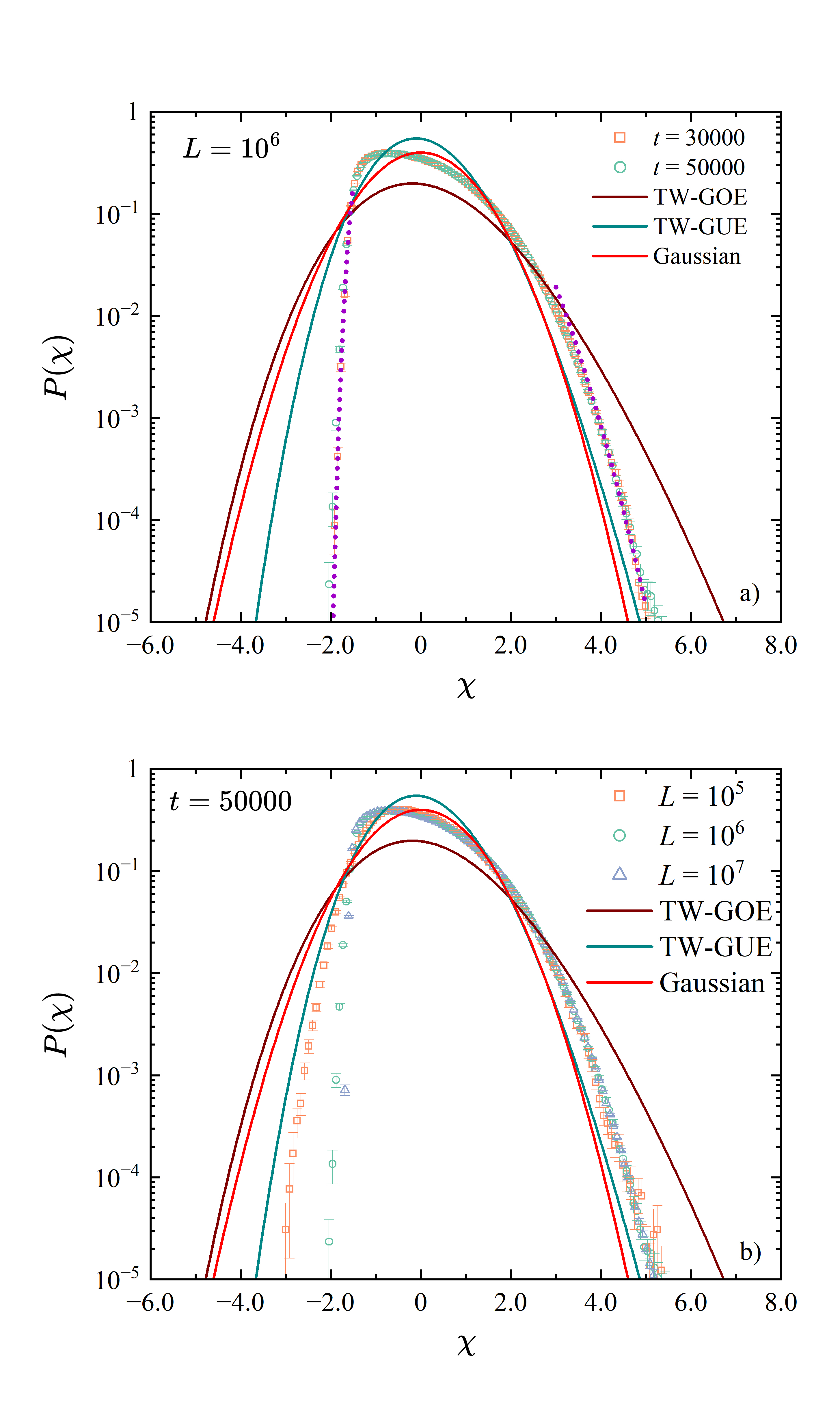}
    \caption{PDF of the front fluctuations for the 1D model. (a) PDF computed for $L = 10^6$ at two different times within the growth regime, see legend. The dotted lines correspond to the best fits of the tails of the PDF to Eq.~\eqref{eq:tails}, with the values for the tail exponents $\eta_{\pm}$ given in Table \ref{tab:SK}. (b) PDF computed at $t = 50000$ for different system sizes, see legend. In both panels, the solid lines show the TW-GOE, TW-GUE, and Gaussian distributions as references.}
    \label{fig:Fluct_1D}
\end{figure}

Figure \ref{fig:Fluct_1D}(a) shows the PDF for the front fluctuations computed for the 1D model at different times within the growth regime, at fixed $L = 10^6$. This plot evidences that, first, once suitably rescaled, the distribution of fluctuations indeed becomes asymptotically independent of time and system size. Second, a markedly asymmetric shape is observed, with a sharp edge for negative $\chi$ and a more regular tail for positive $\chi$. As a reference, we have plotted in Fig.~\ref{fig:Fluct_1D} the Tracy-Widom distributions for the largest eigenvalue of Hermitian random matrices in the Gaussian orthogonal (TW-GOE) and unitary (TW-GUE) ensembles, as well as the familiar Gaussian distribution. The first two are known \cite{Kriecherbauer10,Halpin15,Takeuchi18} to describe the fluctuation PDF in the standard 1D KPZ universality class, in band geometry subject to PBC and for an overall circular symmetry of the front, respectively, whereas the Gaussian distribution holds for linear models of kinetically rough surfaces \cite{Barabasi95, Carrasco19}, all of them in the case of time-dependent (rather than quenched) noise.

Figure \ref{fig:Fluct_1D}(b), on the other hand, displays the PDF for fluctuations computed at a fixed time $t = 50000$ for different system sizes, evidencing stronger size dependence for the left tail of the distribution, whereas the positive tail exhibits an accurate collapse for all values of $L$ considered in this work.  

\begin{figure}[!t]
    \centering
    \includegraphics[width=\columnwidth]{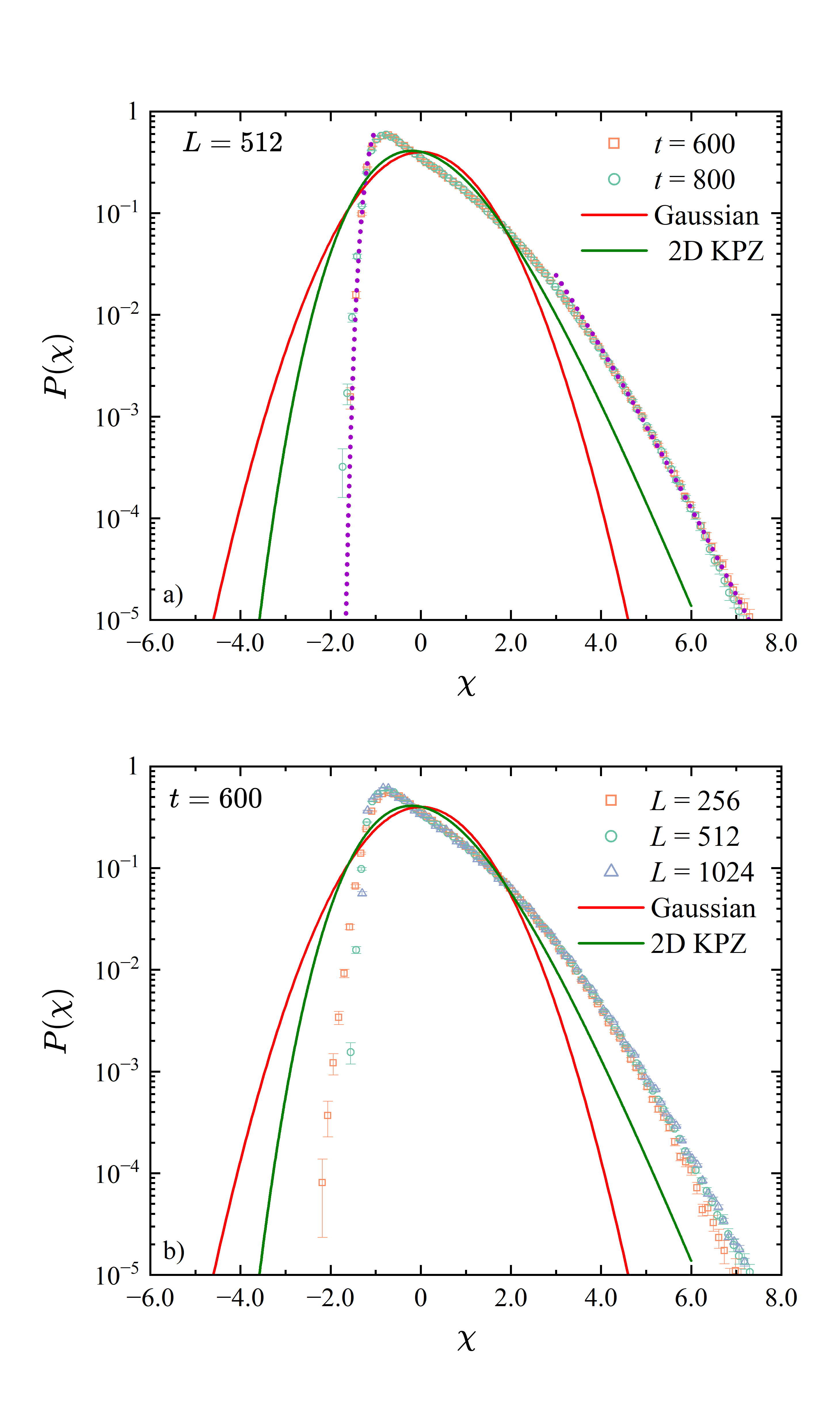}
    \caption{PDF of the front fluctuations for the 2D model. (a) PDF computed for $L = 512$ at two different times within the growth regime, see legend. The dotted lines correspond to the best fits of the tails of the PDF to Eq.~\eqref{eq:tails}, with the values for the tail exponents $\eta_{\pm}$ given in Table \ref{tab:SK}. (b) PDF computed at $t = 600$ for different system sizes, see legend. In both panels and as references, the solid lines show the Gaussian distribution and the Gumbel's first asymptotic form with parameter $m=6$ that describes quite accurately the fluctuation PDF for the 2D KPZ universality class \cite{Oliveira2013} (green).}
    \label{fig:Fluct_2D}
\end{figure}

\begin{figure}[!t]
    \centering
    \includegraphics[width=\columnwidth]{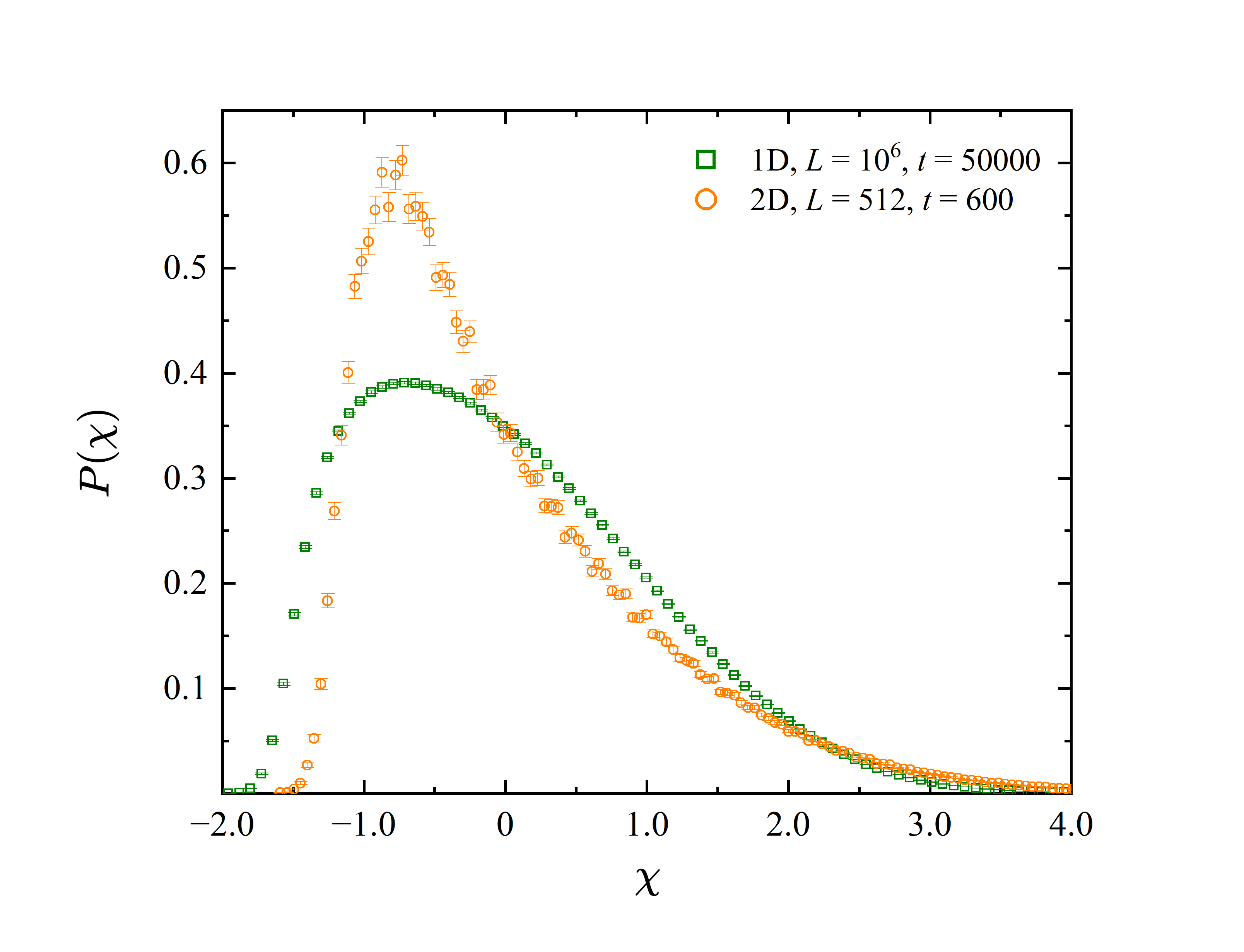}
    \caption{$P(\chi)$ vs.\ $\chi$ plots for both dimensionalities, at selected sizes and times as specified in the legend.}
    \label{fig:Fluct_1D-2D}
\end{figure}

Similar qualitative features are found for $d = 2$, as shown in Fig.~\ref{fig:Fluct_2D}, where the PDFs exhibit quite different shapes, however. This is evidenced in Fig.~\ref{fig:Fluct_1D-2D}, where $P(\chi)$ is plotted as a function of $\chi$ for both dimensionalities, with the data being available
at the Zenodo open access repository \cite{datasetchi}. Note that the PDF is known not to be in the TW family for the 2D KPZ universality class, despite being strongly non-Gaussian \cite{Halpin-Healy2012,Halpin-Healy2013,Oliveira2013}. 

For comparison purposes, we have included in
Fig. 8 the Gaussian distribution together with the so-called
Gumbel’s first asymptotic distribution, as parametrized by
Olivera et al. \cite{Oliveira2013} which accurately describes the fluctuation properties of the 2D KPZ equation (with thermal noise).
Notice that the fluctuations computed for the 2D qKPZ equation clearly differs from that of the thermal KPZ and are not Gaussian. 

The shape parameters, skewness, and excess kurtosis,\footnote{We recall the definition of the skewness $S$ and the excess kurtosis $K$:
$$
S=\langle \chi^3 \rangle_c/\langle \chi^2\rangle_c^{3/2}\,,\,\,\,K=\langle \chi^4 \rangle_c/\langle \chi^2\rangle_c^2 - 3\,,
$$
where $\langle \chi^n\rangle_c \equiv \langle (\chi-\langle \chi \rangle)^n \rangle$.} for the PDFs in Fig.~\ref{fig:Fluct_1D-2D} are given in Table \ref{tab:SK}. Except for the remarkably small value of $K$ for $d=1$, all values imply strong non-Gaussian behavior, with the large skewness accounting for the irreversible motion of the fronts, in spite of the $F=F_c$ conditions implying a vanishing average front velocity.

\begin{table}[t]
\begin{ruledtabular}
\begin{tabular}{cccccc}

$d$ & $L$ & $S$ & $K$ & $\eta_-$ & $\eta_+$ \\

\hline
 1 & $L=10^6$ & 0.6449(9) & $-0.014(3)$ & 7.31(14) & 2.03(4) \\
 2 & $L=512$ & 1.275(5) & 1.84(2) & 6.7(2) & 1.277(7) 
\end{tabular}
\end{ruledtabular}
\caption{Skewness $S$, excess kurtosis $K$, and tail exponents $\eta_{\pm}$ for the front fluctuation distributions in physical dimensions $d=1$ and 2, as obtained in our simulations for the indicated value of $L$ in each case.}
\label{tab:SK}
\end{table}

We consider now the tails of the PDFs in some further detail. As noted from the values of skewness and kurtosis, these tails are asymmetric and non-Gaussian. For many random systems in their disorder-dominated phases, they are frequently seen to follow exponential functions as \cite{Monthus2008}
\begin{equation}
    P(\chi) \approx e^{-c_{\pm}|\chi|^{\eta_{\pm}}}, \quad \chi \rightarrow \pm\infty ,
    \label{eq:tails}
\end{equation}
where $c_{\pm}$ are constants and $\eta_{\pm}$ are characteristic, so-called tail exponents \cite{Monthus2008,Halpin15}. We have fitted the tails of the computed PDF to functions with the form given by Eq.\ \eqref{eq:tails} [dotted lines in Figs.~\ref{fig:Fluct_1D}(a) and \ref{fig:Fluct_2D}(a)] to estimate the $\eta_{\pm}$ exponents; the results are collected in Table \ref{tab:SK} as well. In relation to this, the tail exponents for the TW distribution have been characterized \cite{majumdar2014,Halpin15}. For the $d=1$ KPZ universality class with time-dependent noise, the exponent of the left tail is $\eta_-=3$, while that of the so-called Airy tail (the right tail for TW-GOE in Fig.\ \ref{fig:Fluct_1D}) is $\eta_+=3/2$ \cite{kim1991, majumdar2014}. Moreover, they are related to each other for KPZ as $\eta_-=(d+1)\eta_+$, and to the growth exponent as $\eta_+=1/(1-\beta)$ \cite{majumdar2014,Halpin15}. These relations do not hold for the $\eta_{\pm}$ and $\beta$ values collected in Tables \ref{tab:exponents1D}, \ref{tab:exponents2D}, and \ref{tab:SK}, although the fits are reasonable and such that $\eta_->\eta_+>1$ for $d=1$ and 2, as implied by the KPZ formulae. Violations of the $\eta_+=1/(1-\beta)$ relation have been also reported, e.g., for spreading of precursor films in band geometry \cite{Marcos22} or for synchronized oscillator lattices \cite{Gutierrez2023} (for both of which the PDF is TW-GOE although the kinetic roughening exponents are not 1D KPZ), or for moving fronts associated with the contact process \cite{Barreales2023}.

\section{Discussion}
\label{sec:disc}

\begin{table*}[t]
\begin{ruledtabular}
\begin{tabular}{lcccccc}

$d$ & Model & $\theta$ & $\delta$ & $z$ & $\beta$ & $\alpha$  \\

\hline
\multirow{9}{*}{1} & DPD \cite{Amaral95} & 0.58(7) & $\text{---}$ & $1.01(2)$ & $0.63(1)$ & $0.63(1)$ \\
 & DPD \cite{Tang92, Buldyrev92} & $\approx 0.636$ & $\text{---}$ & $\approx 1$ & $\approx 0.633$ & $\approx 0.633$ \\
 & DPD \cite{Sneppen92} & $\text{---}$ & $\text{---}$ & $\text{---}$ & $\text{---}$ & $\approx 0.63$ \\
 & qKPZ (eq.) \cite{Lee05} & 0.616(9) & 0.360(1) & $\approx 0.978$ & 0.647(2) & 0.633(8) \\
 & qKPZ (DM) \cite{Song07} & $\approx 0.672$ & $\approx 0.377$ & $\approx 1.02$ & $\approx 0.617$ & $\approx 0.627$ \\
 & aDEP \cite{Mukerjee2023} & $\text{---}$ & $\text{---}$ & $\text{---}$ & $\text{---}$ & 0.635(6) \\
 & TL92 \cite{Mukerjee2023} & $\text{---}$ & $\text{---}$ & 1.10(2) & $\text{---}$ & 0.636(8) \\
 & qKPZ (eq.) \cite{Mukerjee2023} & $\text{---}$ & $\text{---}$ & $\text{---}$ & $\text{---}$ & 0.64(2) \\
 & qKPZ (DM) (this work) & 0.673(6) & 0.362(1) & 1.016(7) & 0.6290(6) & 0.6325(17)\\ \hline
\multirow{8}{*}{2} & DPD \cite{Amaral95} & 0.8(2) & $\text{---}$ & 1.15(5) & 0.41(5) & 0.48(3) \\
& DPD \cite{Buldyrev93} & $\text{---}$ & $\text{---}$ & $\approx 1.16$ & $\approx 0.41$ & $\approx 0.48$ \\
 & aDEP \cite{Mukerjee2023} & $\text{---}$ & $\text{---}$ & $\text{---}$ & $\text{---}$ & 0.48(2) \\
 & TL92 \cite{Mukerjee2023} & $\text{---}$ & $\text{---}$ & $\text{---}$ & $\text{---}$ & 0.47(3) \\
 & qKPZ (eq., $F = F_c$) \cite{Wu24} & 1.03(2) & 0.64(2) & 1.32(3) & 0.38(5) & 0.50(3) \\
  & qKPZ (eq., $\lambda > \lambda_c$) \cite{Wu24} & 0.83(2) & 0.58(2) & 1.35(5) & 0.41(1) & 0.51(1) \\
 & qKPZ (DM) (this work) & 0.803(2) & 0.560(3) & 1.182(4) & 0.417(2) & 0.4907(16)\\
\end{tabular}
\end{ruledtabular}
\caption{Comparison of the kinetic roughening exponent values reported for some models in the qKPZ universality class at depinning for $d=1$ and 2 with those obtained in this work. The label ``eq.'' indicates that the results have been obtained by numerically solving the qKPZ equation, whereas DM means discrete model. The ``$\approx$'' symbol indicates that the error bar is not reported or that the exponent is not directly computed. 
}
\label{tab:exponents_previos_b}
\end{table*}

Table \ref{tab:exponents_previos_b} collects our values for the scaling exponents relevant to the kinetic roughening behavior of the qKPZ universality class at depinning, for the two physical dimensions. The table allows for direct comparison with values previously reported in the literature from simulations of several different models in the same universality class ---specifically, the discrete DPD model \cite{Amaral95}, the anisotropic depinning (aDEP) model \cite{Rosso2001,Mukerjee2023}, a cellular automaton proposed in Ref.\ \cite{Tang92} (TL92) and recently reconsidered \cite{Mukerjee2023}, as well as direct simulations of the qKPZ equation \cite{Lee05, Wu24}---, which provide both well-known results and recent revisions thereof. At first sight, our results seem fairly consistent (albeit with reduced statistical errors) with those of the DPD model, and also with additional estimates; actually, the different sets of exponents are compatible with each other. 

In particular, our calculated values for $\alpha$ agree reasonably well with available experimental data for fluid imbibition in paper, which yield $\alpha_{\rm exp} = 0.63(4)$ and $0.52(4)$ at depinning for $d = 1$ and $d = 2$, respectively \cite{Amaral95}. More recent experiments obtain $\alpha_{\rm exp}= 0.66 \pm 0.04$ and $\beta_{\rm exp}= 0.61 \pm 0.05$ for $d=1$ reaction fronts in disordered flows \cite{Atis2015}.

As we have already mentioned, one of the key features of our work is that we have been able to compute the entire set of scaling exponents directly from simulations, i.e., without employing any scaling relation \textit{a priori}. This approach contrasts with that followed by previous authors. In their model for DPD, Amaral {\em et al.}~\cite{Amaral95} computed directly $\beta$ as reported in Table~\ref{tab:exponents_previos_b} (which are virtually identical to ours), but used scaling arguments for the longitudinal and transverse correlation lengths to estimate $\alpha$ for $d \le 6$. For $z$ and $\theta$, they use again scaling arguments (see, for instance, Eqs.~(3.9) and (3.13) in Ref.~\cite{Amaral95}), which may be the reason why their values for $\alpha$ and $z$ deviate more (by up to $\approx -3\%$ for $d = 2$) from ours. This deviation is even larger for $\theta$ in $d = 1$, for which their estimation is roughly 15\% smaller than our calculation. On the other hand, Song and Kim \cite{Song07} used an automaton model in $d = 1$ to find $\beta \approx 0.617$ and $\alpha \approx 0.627$, and estimated $z = \alpha/\beta \approx 1.02$, about 2\% higher than our calculated value. Similarly, Lee and Kim \cite{Lee05} compute $\beta =0.647(2)$ and $\alpha = 0.633(8)$ for the qKPZ equation in $d = 1$ and estimate $z = \alpha/\beta \approx 0.978$, which is around 4\% smaller than our value, the difference arising mostly from our dissimilar $\alpha$ values. Note that, according to our simulations, the $\alpha = \beta z$ and $\delta + \beta = 1$ scaling relations do not hold exactly in the DM, especially for $d = 2$. Given that scaling relations tend to be more accurately satisfied as $L$ increases, we attribute the small deviations found to residual finite-size effects.

Some further discrepancies may be found in Table~\ref{tab:exponents_previos_b}. In Ref.~\cite{Mukerjee2023}, $z=1.10(2)$ was reported for the TL92 model, which is at 4.4 standard deviations away from our estimate and 5 standard deviations from the $z=1$ value conjectured in Ref.\ \cite{Amaral95}. We remark that we have obtained our estimate for $z$ from an analysis of the correlation length extracted from the $C_2(\rr,t)$ correlation function; this analysis yields a value near $z=1$ for all the simulated lattice sizes, see Table \ref{tab:exponents1D}. Besides, our final value, $z=1.016(7)$, is within two standard deviation of the $z=1$ conjectured DPD value. Obviously, we cannot prove the latter using numerical simulations; we can only disprove it, with a given statistical confidence level. 

On the other hand, Wu {\em et al.}~\cite{Wu24} computed the scaling exponents for the 2D model from the numerical simulation of the qKPZ equation. 
These authors address two different conditions, namely $F = F_c$ for $\lambda = 5$ and $F = 0$ at $\lambda = \lambda_c = 2.865$,
which may seem inconsistent with the conditions used in our present work. In this respect, note that a pinning-depinning transition can arise for the qKPZ equation with $F = 0$, provided $\lambda > \lambda_c$ \cite{Ramasco01,Wu24}. In this case, the system behaves as if it were subject to an effective driving force \cite{Ramasco01}, so that our data and those from Wu {\em et al.}~can be properly compared. Actually, Wu {\em et al.}~\cite{Wu24} themselves suggest that the exponents they obtain for $\lambda=\lambda_c$ are ``between those of qEW and qKPZ universality classes to some extent, and belong to the DPD class". Considering both our present results and previous ones (see Table \ref{tab:exponents_previos_b}), the $\lambda = \lambda_c$ case of Ref.\ \cite{Wu24} agrees well with the established values (with their $z$ exponent value being three standard deviations off our estimate). By contrast, their $F=F_c$ case shows noticeable discrepancies, most prominently in the $\delta$ exponent and, to a lesser degree, in the $z$ exponent.

Finally, regarding the PDF of the front fluctuations, we remark that, to our knowledge, no previous analytical or numerical results exist for the qKPZ universality class at depinning. Our results can be naturally compared with those obtained for the standard KPZ universality class (with time-dependent noise), specifically with the TW-GOE distribution for $d=1$ \cite{Kriecherbauer10,Halpin15,Takeuchi18} or, in the case of systems with absorbing states, with the PDF obtained \cite{Barreales2023} for the dynamics of fronts associated with the contact process \cite{Dickman2000}. In both cases the shape of the distributions and the values of skewness and kurtosis differ unambiguously from those we obtain for the qKPZ universality class, in the two space dimensions we have investigated. We expect that the PDF we are presently reporting can also be found at least in other models of the qKPZ universality class at depinning. For the sake of accessibility, we provide our numerical distributions $P(\chi)$ for $d=1$ and 2 in an open-access repository \cite{datasetchi}.

\section{Summary and conclusions}
\label{sec:concl}

In this work, we have simulated numerically the one- and two-dimensional automata versions of the qKPZ equation at the depinning transition. The main findings of the paper, for both dimensionalities, may be summarized as follows:
\begin{itemize}
    \item We have computed the critical force for depinning $F_c$ and the velocity exponent $\theta$, as well as the scaling exponent $\delta$ which rules the dependence of the front velocity with time at $F = F_c$. To our knowledge, the value of $F_c$ for 2D is reported here for the first time.
    \item We have also computed directly the $\alpha$, $\beta$, and $z$
critical exponents associated with the kinetic roughening properties of the fronts. Our numerical values are compatible with (while substantially increasing the accuracy of) previous estimates in the DPD universality class. 
    \item From the computed exponents, we have verified the scaling relations $\beta+\delta=1$ and $\beta z=\alpha$. In this respect, we remark that the exponents $\alpha$ and $z$ were obtained directly from the data, without resorting to scaling relations. Actually, to the best of our knowledge, this is the first time that the entire set of critical exponents has been computed directly from simulations in this system for both dimensionalities.
    \item We have verified the FV scaling of the $C_2(\rr, t)$ correlation function in terms of $\alpha$; no evidences of anomalous scaling have been found.
    \item Finally, we have assessed the occurrence of asymptotically time and system-size independent PDF for the front fluctuations in the growth regime. We have described the shape of these distributions in terms of their skewness, excess kurtosis, and their tail exponents. Specifically, the tails of these PDFs were found to differ noticeably from those commonly found in the disorder-dominated phases of other random systems. Again, this study has been performed for the first time both in 1D and 2D, to our knowledge.
\end{itemize}
    
    Overall, the results reported herein show that the automaton versions describe reasonably well the critical behavior of the qKPZ equation without exhibiting convergence problems or instabilities of the integration schemes.

\section{Acknowledgements}

This work has been partially supported by Ministerio de Ciencia e Innovaci\'on (Spain), by Agencia Estatal de Investigaci\'on (AEI, Spain, 10.13039/501100011033), and by European Regional Development Fund (ERDF, A way of making Europe) through Grant No.\ PID2021-123969NB-I00.
The authors also acknowledge financial support through Grants No.\ PID2024-156352NB-I00 and No.\ PID2024-159024NB-C21, funded by MCIU/AEI/10.13039/501100011033/FEDER, UE and
from Grant No.\ GR24022 funded by the Junta de
Extremadura (Spain) and by European Regional Development Fund (ERDF) “A way of making Europe”, as well as support through grant PIPF-2022/TEC-25428 funded by Comunidad de Madrid (Spain). We have performed
the numerical simulations in the computing facilities of the Instituto de Computaci\'on
Cient\'{\i}fica Avanzada de Extremadura (ICCAEx).

\appendix

\section{Simulation details}\label{appen:details}
The parameters employed in our simulations, namely system sizes, maximum time $t_\text{max}$ achieved and number of realizations of the noise $N_\text{runs}$ are listed in Table \ref{tab:details} for the 1D and 2D models. The maximum times were chosen in such a way that the system had entered the pinned phase under each condition (the only exceptions are the computation of $C_2(\rr,t)$ for $L=10^6$ and $w(t)$ for $L=10^6$ and $L=10^7$ in the 1D model). All simulations were run at $F = F_c$ for each dimensionality.  

\begin{table}[!h]
\begin{ruledtabular}
\begin{tabular}{cccc}
 $d$ & $L$ & $t_{\text{max}}$ & $N_\text{runs}$ \\
\hline
\multirow{6}{*}{1} & $10^4$ & $5 \times 10^5$; $10^4$ & 4000; 5000  \\
  & $5\times 10^4$ &  $5\times 10^5$; $10^5$  & 2000; 4000  \\
  & $10^5$ &  $10^6$ ; $10^5$  & 4000; 2000  \\
  & $5\times 10^5$ & $7.7 \times 10^6$;   $5 \times 10^6$ & 3000; 1000  \\
  & $10^6$ & $10^6$; $1.5\times 10^5$& 5000; 1152 \\
  & $10^7$ & $5 \times 10^5$; \text{---} & 2000 ; \text{---}\\
\hline
\multirow{5}{*}{2} & 128 & 4 $\times 10^3$ & 4000 \\
  & 256 & 12 $\times 10^3$ & 4000 \\
  & 512 & 23 $\times 10^3$ & 4000 \\
  & 1024 & 53 $\times 10^3$ & 4000 \\
  & 2048 & 72 $\times 10^3$ & 4000 \\

\end{tabular}
\caption{Parameter values for our numerical simulations. Here, $t_\text{max}$ is the number of updates of the front (1 update means the actualization of all the $N=L^d$ heights of the front) and $N_\text{runs}$ is the number of noise realizations simulated. For $d=1$ we have performed two sets of simulations for each $L$: one for the measurement of the roughness and average position of the front and a second one for the computation of the full $C_2(\rr,t)$ correlation function. For this reason, we report two sets of numbers in the columns for $t_\text{max}$ and $N_\text{runs}$. In addition, to compute the $\alpha$ exponent in the 1D model we have simulated $L=100$ (2000 runs) and $L=1000$ (4000 runs). Finally, we have simulated $L=10^7$ (100 runs) with $t_\text{max}=10^5$ for the computation of the $z$ and $\alpha$ exponents in this size.}
\label{tab:details}
\end{ruledtabular}
\end{table}

\section{Correlation function}
\label{appen:scaling}
\begin{figure}[!b]
    \centering
    \includegraphics[width=\columnwidth]{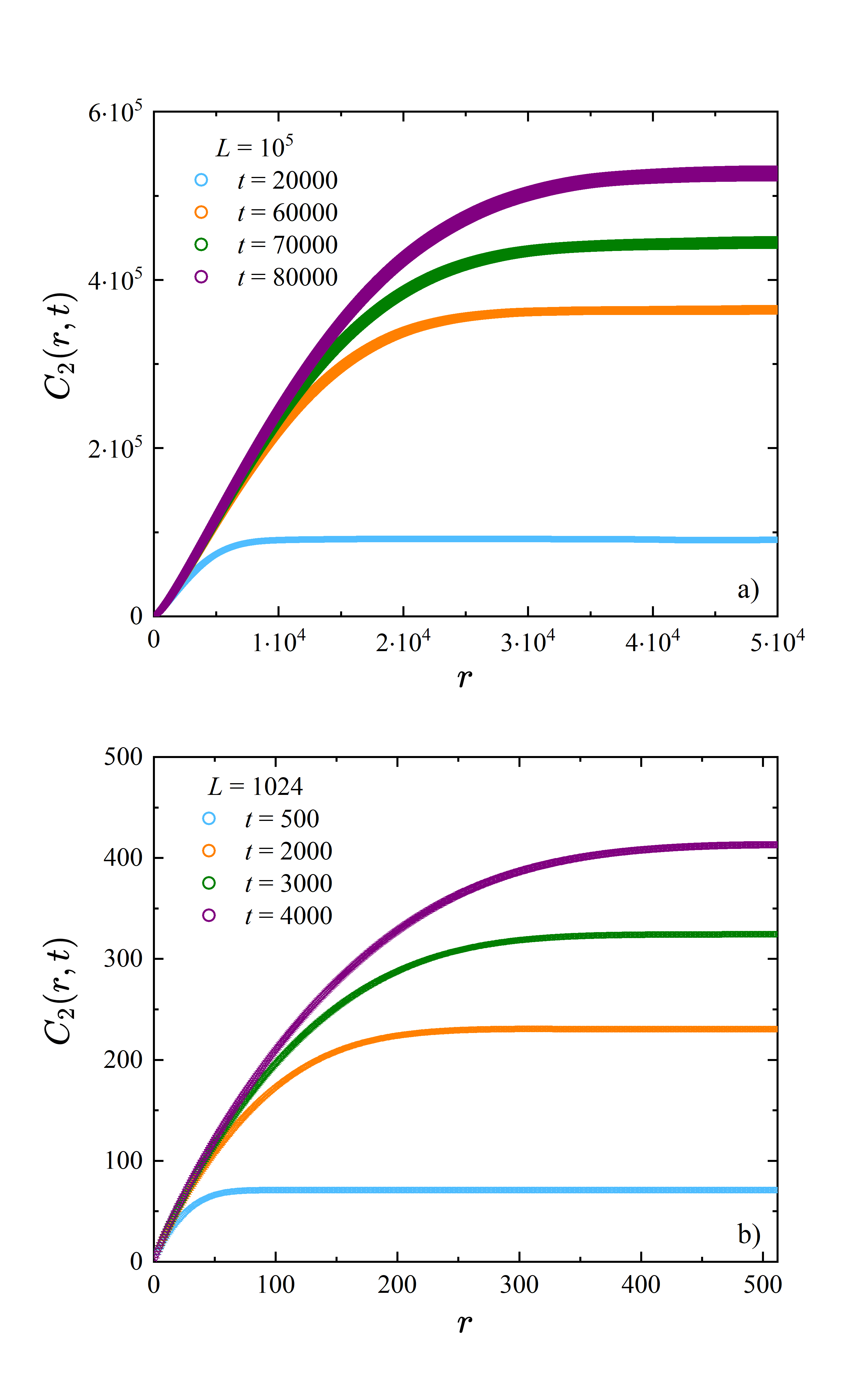}
    \caption{$C_2(\rr, t)$ vs.\ $r$ for several times as given in the legend, for (a) $d=1$ with $L = 10^5$ and (b) $d=2$ with $L = 1024$.}
    \label{fig:C2_unscaled}
\end{figure}

Figure \ref{fig:C2_unscaled} shows two sample unscaled $C_2(\rr,t)$ functions, arbitrarily taken as representatives for both the 1D (a) and 2D (b) models, for several times, and fixed $L$. As previously mentioned, no signs of anomalous scaling (typically, a time-dependent shift of the $C_2(\rr,t)$ vs.\ $r$ curves for increasing $t$, see e.g.\ Ref.~\cite{Barreales2023}) are observed, which is a sound demonstration of the validity of the FV scaling for both dimensionalities. 

%\bibliography{Automaton.bib}

\end{document}